\documentstyle[12pt]{article}
\begin{document}
\begin{titlepage}
\vspace*{-62pt}
\begin{flushright}
{\small
FERMILAB--Pub--95-091-A \\
April 1995 \\
}
\end{flushright}
\vspace{1in}
\begin{center}
\Large
{\bf Decaying $\Lambda$ cosmologies and statistical properties of
gravitational lenses}
\vspace{.35in}
\normalsize
 L. F. Bloomfield Torres$^{\diamond ,}$ and I. Waga$^{\ast ,
\diamond}$
\normalsize
\vspace{.7cm}
{\em $^{\ast}$NASA/Fermilab Astrophysics Center,  \\
Fermi National Accelerator Laboratory, Batavia, IL 60510}
\vspace{.35cm}
{\em $^{\diamond}$Universidade Federal do Rio de Janeiro, Instituto
de
F\'isica, \\
Rio de Janeiro - RJ - Brasil -21943}
\end{center}
\vspace{.7in}
\baselineskip=24pt
\begin{abstract}
\noindent

In this paper we investigate the statistical properties of
gravitational lenses
for models in which a cosmological term decreases with time as
$\Lambda
\propto a^{-m}$, where $a$ is the scale factor and $m$ is a parameter
($0 \leq
m < 3$). We show that for given low values of the present matter
density
parameter $\Omega_{m0}$, there is a wide range of values for $m$ for
which the
lensing rate is significantly smaller than that in cosmological
constant
($\Lambda$) models. We also show that models with low $\Omega_{m0}$
and
$m\stackrel{>}{\sim}2$ have high likelihood to reproduce the observed
lens
statistics in the HST snapshot survey.
\end{abstract}
\normalsize
\end{titlepage}
\section{Introduction}
\noindent

In the last five years or so, the statistics of gravitational lensing
(Turner, Ostriker and Gott III 1984, hereafter TOG) has proven to be
a powerful
tool in
constraining models of the universe, especially those with a
cosmological
constant ($\Lambda $). Cosmologies with a $\Lambda$-term have a long
history
and are now receiving considerable attention (see Carroll, Press
and Turner 1992 for review). Spatially flat cosmological models with
a
cosmological constant have been suggested (Peebles 1984; Turner,
Steigman
\& Krauss 1984) as a way to reconcile inflation with dynamical
analyses on scales of $\sim 10h^{-1}$ Mpc, that indicate a value for
the
density parameter $\Omega _0\sim 0.1$ to $0.3$ (Peebles 1993). A
cosmological constant also alters the transfer function for the
density
perturbations giving more power in
the perturbation spectrum at large scales (as compared with standard
CDM) in
accordance with observations (Efstathiou {\it et al.} 1990; Lahav
{\it et al.}
1991; Kofman {\it et al.} 1993). Besides, if the present value of the
Hubble
parameter is high, as indicated by some recent observations (Pierce
{\it et.
al.} 1994; Freedman {\it et. al.} 1994), a cosmological term will be
the
only way to get a theoretical age for a flat universe in accordance
with
current age estimates for globular clusters (Chaboyer 1994).

The idea that light could be focused by the gravitational lens effect
was
first suggested by Lodge (1919) near the beginning of the century.
For several
decades the subject of gravitational lensing had a quite slow
development,
but recently it started to become one of the most active research
area in
astrophysics and cosmology. There are several reasons for the current
interest in gravitational lensing. On the cosmological side, after
the works
of Refsdal (1964) and Press \& Gunn (1973), it was realized that
cosmological parameters could be probed by the gravitational lensing
effect.
In the beginning it seemed that lensing properties were too
insensitive and
would only distinguish extreme cosmological models. Later on, Turner
(1990)
and Fukugita, Futamase \& Kasai (1990) showed that a non-zero
cosmological
constant could significantly affect the statistics of gravitational
lenses,
especially in a low-density universe.

However, there are uncertainties in the study of the statistics of
gravitational lensing (Mao 1991; Fukugita, Futamase, Kasai \& Turner
1992,
hereafter FFKT). For example, the lens effect depends considerably on
how the
mass is distributed in the lensing galaxy. Hinshaw and Krauss (1987)
showed
that the introduction of a core in the isothermal sphere galaxy model
(non-singular isothermal sphere) can significantly modify the
statistical
lensing properties. Another issue is what distance formula should be
used: the
angular diameter distance or the Dyer-Roeder distance? Related to
this question
is the kind of statistics to be applied (Ehlers \& Schneider 1986;
FFKT). We
should also mention the important effect of magnification bias and
other
selection effects such as angular resolution, galaxy evolution and
merging on
lensing probabilities (TOG; Fukugita \& Turner 1991, hereafter FT;
Mao 1991;
Mao \& Kochaneck 1993; Rix, Maoz, Turner \& Fukugita 1994).

In spite of the uncertainties, the calculated rate of lensing in
$\Lambda$-flat
models, when confronted with the existing lensing observations,
indicates that
models with density parameter ($\Omega_{m0}$) close to unity are most
likely.
For instance, Maoz \& Rix (1993) claim that at present we should have
$\Omega
_\Lambda \stackrel{<}{\sim}0.7 $ . So, it is becoming more and more
difficult
to make the dynamical estimates for $\Omega$ on scales $\sim 10
h^{-1}$ Mpc
compatible with a flat cosmological model with $\Lambda \neq 0$. It
should be
pointed out, however, that the lensing frequency of quasar images is
considerably reduced if
early-type galaxies ($z \stackrel{>}{\sim} 0.5$) were dusty (Fukugita
\&
Peebles (1994)).

The results above seem to favor open FRW models. There are however
other
possibilities. For example, Ratra and Quillen (1992) showed that, for
a wide
range of parameters, the predicted lensing rate is considerably
reduced in a
class of flat models (Peebles \& Ratra 1988, Ratra \& Peebles 1988)
in
which a scalar field plays the role of an effective cosmological
``constant'' that decreases with time. Some other models with a
decreasing
cosmological term were also proposed ((Ozer \& Taha 1987a,b; Freese,
Adams,
Frieman \& Mottola  1987; Chen \& Wu 1990; Abdel-Rahman 1992;
Carvalho, Lima \& Waga 1992; Silveira \& Waga 1994) and it would be
interesting
to know if they also predict a lower lensing rate. We can argue that
we should
expect a positive answer to this question. The reason is that usually
in a
varying $\Lambda$ cosmological model, the distance to an object with
redshift
$z$ is smaller than the distance to the same object in a constant
$\Lambda$
model with the same $\Omega_{m0}$. So, the probability that light
coming from
the  object is affected by a foreground galaxy is reduced in a
decaying
$\Lambda$ cosmology. However this is only a qualitative argument, and
it is
clear that a  quantitative treatment is necessary if we want to put
limits on
parameters of the models.

In this paper we address the above question to the special class of
models
proposed by Silveira and Waga (1994) in which a cosmological term
decreases
with time as $\Lambda \propto a^{-m}$, where $a$ is the scale factor
and $0
\leq m < 3$ is a constant. We show that these models also admit a
large set of
parameters for which the predicted lensing rate is much lower than
that
obtained in a constant $\Lambda$ model with the same low value of
$\Omega_{m0}$.
The paper is organized as follows: In section ($2$) the assumptions
and
basic equations of our models are presented. We exhibit expressions
for two
sorts of distance that we shall use, and discuss the corresponding
statistics associated with them. In section ($3$) we model galaxies
by the
isothermal sphere density profile and obtain the predicted lensing
probabilities and the distribution of image angular separation for
some typical
models previously chosen. In section ($4$) we
compare the predictions of the models to observations and stress our
main
conclusions in section ($5$).

\section{Decaying vacuum cosmological models -- distance and optical
depth
formulas}

\noindent

Following Silveira and Waga we assume that the cosmic fluid is a
non-interacting
mixture of non-relativistic matter and radiation. The cosmological
term is assumed to be a time dependent quantity,
\begin{equation}
\Lambda =3\tilde \beta a^{-m},
\end{equation}
where $\tilde \beta \geq 0$ is a constant and the factor 3 was
inserted for
mathematical convenience. We also assume that the vacuum decays only
into
relativistic particles such that matter is conserved ($\rho _m\propto
a^{-3}$
). As shown by Silveira and Waga (1994), the radiation energy density
has two
parts; one conserved ($\Omega _{r0}{H_0}^2(a_0/a)^4$) and a second
one, ($\frac{3m\tilde \beta }{8\pi G(4-m)}a^{-m}$), that appears due
to
particle creation by the decaying vacuum. Here $a_0$ is the present
value of
the scale factor, $H_0=100$ $h$ $km$ $s^{-1}$ $Mpc^{-1}$ is the
present
value of the Hubble parameter ($h \simeq 0.5 - 1$) and $\Omega
_{r0}=4.3\times 10^{-5}h^{-2}$ stands for the present value of the
conserved
radiation density parameter. In the following, subscripts $0$ will
always
indicate present values.

The Einstein equations for the models we are considering reduce to
\begin{equation}
\left( \frac{\stackrel{.}{a}}a\right) ^2=\Omega _{m0}H_0^2\left(
\frac{a_0}
a\right) ^3+\Omega _{x0}H_0^2\left( \frac{a_0}a\right) ^m -
\Omega_{k0}H_0^2\left( \frac{a_0}a\right) ^2,
\end{equation}
and
\begin{equation}
\frac{\stackrel{..}{a}}a=-\frac 12\Omega _{m0}H_0^2\left(
\frac{a_0}a\right)
^3+\frac{(2-m)}2\Omega _{x0}H_0^2\left( \frac{a_0}a\right) ^m,
\end{equation}
where $\Omega _{m0}$ is the matter density parameter, $\Omega
_{x0}=\frac{
4\tilde \beta H_0^{-2}a_0^{-m}}{(4-m)}$ and $\Omega_{k0} =\frac{k}{
H_{0}^2a_{0}^2}$.

Since we are mainly interested in the lensing properties of the
models, only
recent epochs have to be considered ($z \stackrel{<}{\sim} 5$). This
justifies neglecting the conserved radiation energy density on the
right
hand side of (2) and (3). To have some grounds of comparison we have
included the curvature term in ($2$) and will also consider the open
FRW ($
k=-1$) model.

The equations above are quite general and apply for a broad spectrum
of
models. Let us first consider the $k=0$ case. For instance, if $m=0$
the
usual flat FRW model with a cosmological constant is recovered,
whilst if $
m=2$ the above equations (with $\Omega _{k0}=0$) are formally the
same as
those of the open FRW model. The same equations also appear in some
string
dominated cosmologies (Vilenkin 1984). Further, we would get the same
equations if we had considered, besides matter , an exotic x-fluid
with
equation of state, $p_x=(\frac m3-1)\rho _x$. Cosmologies having a
fluid
with this behavior were investigated by Fry (1985), Sahni, Feldman
and
Stebbins (1992), Feldman and Evrard (1993) and more recently by
Martel (1995).
We remark that all we shall
discuss here also applies for these models. We should also mention
that in
the limit $\rho _m>>\rho _\phi $, the scalar field model analyzed by
Ratra
and Quillen has the same behavior as the one proposed here. This can
be seen
easily if we relate their parameter $\alpha $ with $m$ as, $\alpha
=\frac{2m
}{3-m}$. It is clear however, that the models are different (unless
$m=0$)
if $\Omega _{m0}\stackrel{>}{\sim }\Omega _{x0}$ or during the
x-component
(vacuum) dominated era when, in fact, all the lensing properties we
shall
discuss are important.

In this paper we shall compare the following models:\\ {\bf Case A}:
$m=0$, $
\Omega _{m0}=1$ and $k=0$ (Einstein-de Sitter model).\\ {\bf Case B}:
$m=2$,
$\Omega _{m0}=0.2$ and $k=0$.\\ {\bf Case C}: $m=0$, $\Omega
_{m0}=0.2$ and $
k=0$ (Friedman-Lemaitre model).\\ {\bf Case D}: $\Omega _{m0}=0.2$,
$\Omega
_{x0}=0$ and $k=-1$ (Open FRW model).\\ For the sake of completeness
we have
included case D in our analysis. It will be interesting to compare it
with
case B, which has similar field equations but has flat spatial
sections. We
also analyzed the case $m=1$ and it turned out that it always has
behavior
between cases B and C, so we decided not to explicitly include it in
our
discussion.

For the flat models A, B and C, the angular diameter distance,
$d_S(z_L,z_S)$
, between two objects, one with redshift $z_L$ and the other with
$z_S$
is given by,
\begin{equation}
d_{S}(z_L,z_S)=\frac{cH_0^{-1}}{1+z_S}\int_{z_L}^{z_S}
\frac{dy}{\sqrt{
\Omega_{m0}(1+y)^3+(1-\Omega_{m0})(1+y)^m}}.
\end{equation}
Equation ($4$) can be expressed in terms of the hypergeometric
functions $
F(a,b;c,z)$ as
\begin{eqnarray}
d_{S}(z_L,z_S ) &=& \frac{2cH_0^{-1}}{(1+z_S)
\sqrt{\Omega_{m0}}}\times \nonumber \\
& &(\frac{1}{\sqrt{1+z_L}}
F(\frac12,\frac{1}{6-2m};\frac{7-2m}{6-2m},
-\frac{1-\Omega_{mo}}{\Omega_{m0}}(1+z_L)^{m-3})-  \nonumber \\
& &\frac{1}{\sqrt{1+z_S}}F(\frac12,\frac{1}{6-2m};\frac{7-2m}{6-2m},
-\frac{1-\Omega_{mo}}{\Omega_{m0}}(1+z_S)^{m-3})).
\end{eqnarray}
For some special values of $m$, the hypergeometric function in ($5$)
can be
reduced to elementary functions. This can be done, for instance, for
$m=2$
if we use the relation, $F(1/2,1/2;3/2,-x^2)=x^{-1} \sinh^{-1} (x)$.
For
the Einstein-de Sitter model equation ($4$) can easily be integrated
giving
\begin{equation}
d_{S}(z_L,z_S)=\frac{2c{H_0}^{-1}}{1+z_S}\left[(1+z_L)^{-1/2}
-(1+z_S)^{-1/2}\right].
\end{equation}
In fact, we can obtain ($6$) from ($5$), by observing that in the
limit $
\Omega_{m0}\rightarrow 1$ the hypergeometric function also goes to
unity.

In the case of open models, the angular diameter distance can be
expressed
as (FFKT)
\begin{eqnarray}
d_{S}(z_L,z_S )&=&\frac{2cH_0^{-1}}{\Omega_{m0}^2(1+z_L)(1+z_S)^2}
\{(2-\Omega_{m0}+\Omega_{m0}z_S)\times \nonumber \\
&&\sqrt{1+\Omega_{m0}z_L}-
(2-\Omega_{m0}+\Omega_{m0}z_L)\sqrt{1+\Omega_{m0}z_S}\}.
\end{eqnarray}

The differential probability, $d\tau$, that a line of sight
intersects a
galaxy at redshift $z_L$ in the interval $dz_L$ from a population
with
number density $n_l$ is (TOG; Peebles 1993)
\begin{equation}
d\tau = - n_l \sigma c \frac{dt}{dz_L}dz_L ,
\end{equation}
where from ($2$) we have
\begin{equation}
\frac{dt}{dz}=-\frac{H^{-1}(z)}{1+z} =-\frac{H_0^{-1}}{1+z}
\left(\Omega_{m0}\left( 1+z \right) ^3+\Omega _{x0}\left( 1+z \right)
^m -
\Omega_{k0}\left( 1+z \right) ^2 \right)^{-1/2}.
\end{equation}
The cross section ($\sigma$) in ($8$) is given by
\begin{equation}
\sigma= \pi a_{cr}^2,
\end{equation}
where $a_{cr}$ is the effective radius of the lens, that is, $a_{cr}$
is the
maximum distance of the lens from the optical axes for which multiple
image
is possible.

The total optical depth ($\tau$) is obtained by integrating $d\tau$
along
the line of sight from $0$ to $z_S$, that is
\begin{equation}
\tau = \int_0^{z_S}d\tau = -\int_0^{z_S} n_l \sigma c
\frac{dt}{dz_L}dz_L .
\end{equation}

In the angular diameter distance definition it is assumed that the
matter in
the universe is homogeneously distributed. However the gravitational
lens
effect will not occur in a smooth universe. Only if matter is
clumped, as in
the real universe, can this effect take place. A distance formula
that takes
matter clumping into account was proposed by Dyer and Roeder (1972,
1973)
and is known as the Dyer-Roeder (DR) distance. Here we will consider
two
extreme cases. We have already discussed the first one in which the
smoothness parameter ($\tilde \alpha$), where $0\leq \tilde \alpha
\leq1$,
is equal to one (filled beam DR distance or angular diameter
distance). The
other extreme case, $\tilde
\alpha =0$, is called the DR empty beam distance (Schneider, Ehlers
and
Falco 1992).

For the models under consideration, the empty beam DR distance is
given by (
FFKT)
\begin{eqnarray}
d_{DR}(z_L,z_S)&=& cH_0^{-1}(1+z_L) \times \nonumber \\
&&\int_{z_L}^{z_S}\frac{(1+y)^{-2}dy}{\sqrt{\Omega_{m0}(1+y)^3+(1-
\Omega_{m0}-\Omega_{x0})(1+y)^2+\Omega_{x0}(1+y)^m}}.
\end{eqnarray}
Notice that for the same $\Omega_{m0}$, flat models ($\Omega_{m0}
+\Omega_{x0}=1$) with $m=2$ and open models ($\Omega_{x0}=0$) have
the same
empty beam distance. For the open ( just make $m=2$ in ($13$) ) and
flat
models, equation ($12$) can be expressed in terms of hypergeometric
functions as
\newpage
\begin{eqnarray}
d_{DR}(z_L,z_S)&=&
\frac{2cH_0^{-1}(1+z_L)}{5\sqrt{\Omega_{m0}}}\times \nonumber \\
&&((1+z_L)^{-5/2}F(\frac12,\frac{5}{6-2m};\frac{11-2m}{6-2m},
-\frac{1-\Omega_{mo}}{\Omega_{m0}}(1+z_L)^{m-3})-  \nonumber \\
& &(1+z_S)^{-5/2}F(\frac12,\frac{5}{6-2m};\frac{11-2m}{6-2m},
-\frac{1-\Omega_{mo}}{\Omega_{m0}}(1+z_S)^{m-3})).
\end{eqnarray}
Again, in the limit $\Omega_{m0}\rightarrow 1$ the hypergeometric
function
goes to unity and ($13$) simplifies to
\begin{equation}
d_{DR}(z_L,z_S)=\frac{2cH_0^{-1}(1+z_L)}{5\sqrt{\Omega_{m0}}} (
(1+z_L)^{-5/2}- (1+z_S)^{-5/2}).
\end{equation}

In obtaining the probability of multiple images in ($11$), we
considered a
random line of sight to the source at $z_S$, calculated the expected
number
of lenses ($d\tau$) in the redshift interval $dz_L$ around $z_L$, and
then
integrated $d\tau$ from $0$ to $z_S$. Ehlers and Schneider (1986)
observed
that in a self-consistent treatment of probabilities in a clumpy
universe,
the random variable should be the position of the source on a sphere
at $z_S$
(and not the line of sight to the source). They then proposed a new
derivation for the optical depth that is called the ES probability.
The
Ehlers-Schneider differential probability ($d\tau_{ES}$) can be
expressed as
(FFKT)
\begin{equation}
d\tau_{ES}=\left(\frac{d_{DR}(0,z_S)}{d_{S}(0,z_S)}\right)^2
\left(\frac{
d_{S}(0,z_L)}{d_{DR}(0,z_L)}\right)^2 d\tau,
\end{equation}
where in $d\tau$ (given by equation ($8$)), the empty beam distance
should be
used.
By integrating ($15$) from $0$ to $z_S$ we obtain the total ES
optical depth.

In Figures $1a$ and $1b$ we present the quantity $D_{OS}/cH_0^{-1}$
for
the filled and empty beam distances. We also show in Figures $2a$ and
$2b$
, also for both distances, the combination
$D_{OL}D_{LS}/(D_{OS}cH_0^{-1})$
that appears, through $a_{cr}$,  in the differential probability
formulas. We
are following
the TOG and FFKT notation, such that $D_{LS}=d(z_L,z_S)$,
$D_{OS}=d(0,z_S)$
and $D_{OL}=d(0,z_L)$.

\section{The isothermal sphere galaxy model}
\noindent

Let us now consider the isothermal sphere model for
galaxies. This model is characterized by two parameters, namely, the
core
radius ($r_c$) and the one-component velocity dispersion
($\sigma_{||}$).
Following Hinshaw and Krauss (1987), we assume
the lens galaxy to have the following density profile
\begin{equation}
\rho(r)=\frac{\sigma_{||}^2}{2\pi G(r^2+{r_c}^2)}.
\end{equation}
The surface mass density of the lens on the lens plane is given by
(Bourassa \&
 Kantowski 1975)
\begin{equation}
\Sigma(a)=2\int_a^{\infty}\rho(r)\frac{rdr}{\sqrt{r^2-a^2}}= \frac{
\sigma_{||}^2}{2 G \sqrt{a^2+{r_c}^2}},
\end{equation}
and the projected mass interior to the impact parameter $b$ is,
\begin{equation}
M(b)=2\pi\int_0^ba\Sigma(a)da=\frac{\pi \sigma_{||}^2}{G}
\left[\sqrt{b^2+{
r_c}^2}-r_c \right].
\end{equation}
The bending angle is
\begin{equation}
\alpha=\frac{4GM(b)}{c^2b}=\alpha_0 \frac{\sqrt{b^2+{r_c}^2}-r_c}{b},
\end{equation}
where $\alpha_0=4\pi\left(\frac{\sigma_{||}}{c}\right)^2 \approx
1.8^"$$
(\sigma_{||}/250$ $km$$s^{-1})^2$ denotes the constant bending angle
for the
singular isothermal sphere case (SIS), obtained by taking the limit $
r_c\rightarrow 0$.

By using simple geometry it is easy to see from Figure 3 that
\begin{equation}
l+b=\frac{D_{OL}D_{LS}}{D_{OS}} \alpha,
\end{equation}
where $l$ is the distance from the lensing galaxy to the unperturbed
line of
sight. It follows from ($19$) and ($20$) that, if $r_c=0$ (SIS case),
the
maximum value of $l$ for multiple images ($a_{cr}$) is given by
\begin{equation}
a_{cr}(0)=\frac{D_{OL}D_{LS}}{D_{OS}} \alpha_0.
\end{equation}
If $r_c\neq0$ (NSIS), by substituting ($19$) in ($20$) we get the
cubic
equation,
\begin{equation}
b^3 + 2lb^2+(l^2+2r_ca_{cr}(0)-a_{cr}^2(0))b+2lr_ca_{cr}(0)=0.
\end{equation}
The number of real and distinct solutions of ($22$) depends on the
sign of its
discriminant. Hinshaw and Krauss (1987) showed that in order to
produce multiple images, the lens maximum distance from the
unperturbed line
of sight should have the following expression
\begin{equation}
a_{cr}= a_{cr}(0)\left[(1+5\beta-\frac12\beta ^2)-
\frac12\beta^{1/2}(\beta+4)^{3/2}\right]^\frac12,
\end{equation}
where $\beta=r_c/a_{cr}(0)$. Further, from the multiple image diagram
(Young
{\it et. al.} 1980, Blandford and Kochanek 1987), they also showed
that
multiple images are only possible if $\beta<1/2$. So, the cross
section for
NSIS is $\sigma=0$ for $\beta>1/2$ and $\sigma=\pi a_{cr}^2$ (with
$a_{cr}$
given by equation ($23$)) if $\beta<1/2$.
In fact, if  $\beta < 1/2$, instead of two (as in SIS), three images
are
predicted, in agreement with
the odd number of images theorem valid for symmetric (non-singular)
lenses
(Subramanian \& Cowling 1986). It can also be shown (Hinshaw and
Krauss; see
also Hinshaw 1988 for more details) that the angular image separation
between
the two outer images is given by
\begin{equation}
 \theta =\frac{2 a_{cr}(0)}{D_{OL}} \left[ \left(1-\beta \right)^2
-\beta
^2\right] ^{1/2} .
\end{equation}
Notice that if $\beta=0$ the usual result $\theta = 8 \pi
\frac{D_{LS}}{D_{OS}}\left(\frac{\sigma_{||}}{c}\right)^2$, valid for
SIS
lenses, is recovered.

In the SIS case, by using equation ($21$) and assuming conserved
comoving
number density of lenses ( $n_l=n_0(1+z)^3
$ ), equation ($11$), can be analytically integrated. By using
standard
distance and statistics we obtain for flat models,
\begin{equation}
\tau(z_S) = \frac{f}{30}\left(d_S(0,z_S)(1+z_S)\right)^3.
\end{equation}
Here,
\begin{equation}
 f=\frac{16\pi^3}{cH_0^3}<n_0\sigma_{||}^4>
\end{equation}
measures the effectiveness of the lens in producing multiple images
(TOG).
An analytic expression for $\tau$ in the case of filled beam distance
and
standard statistics can also be obtained for $k\neq0$ (see
Gott, Park and Lee 1989).

Following FT, we consider the existence of 3 species of galaxies (E,
SO and S)
and assume a Schechter form for the luminosity function,
\begin{equation}
\Phi(L) dL = \phi^{\star}(\frac{L}{L^{\star}})^\alpha
\exp(-L/L^{\star})\frac{dL}{L^{\star}},
\end{equation}
where $ \phi^{\star}=(1.56 \pm 0.4)\times 10^{-2} h^3$$Mpc^{-3}$
(Efstathiou,
Ellis \& Peterson, 1988) is a galaxy number density and $\alpha =
-1.1 \pm 0.1$
(see FT). The morphological composition is assumed to be
$E:SO:S=12:19:69$.
We assume, in addition, the relationship,
$(\frac{L}{L^{\star}})=(\frac{\sigma_{||}}{{\sigma_{||}}^{\star}})^\gamma$,
between galaxy luminosity and velocity dispersion. Here the exponent
$\gamma$
is, $\gamma=4$ for E/SO ( Faber and Jackson 1976) and $\gamma=2.6$
for S
galaxies (Tully and Fisher 1977). Substitution of  (27) in (26) leads
to,
\begin{equation}
f=\frac{16\pi^3}{cH_0^3} \phi^{\star}{\sigma_{||}^\star}^4
\Gamma(\alpha+\frac4\gamma +1),
\end{equation}
where $\Gamma(x)$ is the Gamma function and $\sigma_{||}^{\star}$ is
the
velocity dispersion corresponding to the characteristic luminosity
$L^{\star}$.

Fukugita and Turner estimated $\sigma_{||}^\star$ to be:
\begin{eqnarray}
\sigma_{||}^\star=
\left\{
\begin{array}{lll}
225_{-20}^{+12} &for\;\; {\rm E},\\
206_{-20}^{+12} &for\;\; {\rm SO},\\
144_{-13}^{+8} &for\;\; {\rm S}.
\end{array}
\right.
\end{eqnarray}
To take into account dark massive halos, they follow TOG and also
adopted a
$(3/2)^{1/2}$ correction factor for the velocity dispersions for E/SO
galaxies.
With the above numbers we find $f_E=0.018\pm 0.009$, $f_{SO}=0.020\pm
0.011$,
and $f_S=0.007\pm 0.003$ (total effectiveness parameter, $f=0.045\pm
0.023$).
More recently Kochanek (1993) argued that the ratio of dark matter
dispersion
velocity to that of luminous matter should be in the range $0.9-1.05$
and
suggested that the $(3/2)^{1/2}$ correction factor should not be
considered.
Without this the factors $f_{E/SO}$ are $2.25$ smaller, that is,
$f_E=0.008\pm
0.004$, $f_{SO}=0.009\pm 0.005$ and we get $f=0.024\pm 0.012$. We
shall
consider in the next section these two cases when comparing the
predictions of
the models with observations.

In Figures $4a$ and $4b$ the normalized optical depth ($\tau / f$),
for the
four models in the SIS case, is displayed for filled beam distance
and standard
statistics (Figure $4a$) and for empty beam distance and
Ehlers-Schneider
statistics (Figure $4b$). We also obtained the optical depth for the
NSIS case.
As discussed before, in this case we should use the appropriate cross
section
with $a_{cr}$ given by ($23$). For an analytic expression of the NSIS
optical
depth in the standard case see Krauss and White (1992). In Figure
$5a$ is
displayed the NSIS normalized optical depth for filled beam distance
(standard
statistics). In Figure $5b$ the same quantity is displayed for the
empty beam
distance (Ehlers-Schneider statistics). A constant value for the core
radius,
$r_c=0.5$$h^{-1} kpc$ and a velocity dispersion
$\sigma_{||}^\star=144$ $km/s$
are assumed for all the models.

The present available data do not allow very good estimates for the
core radius
of galaxies. In fact, $r_c$ seems to vary a lot even among galaxies
with the
same morphology. However there is some evidence that most spirals
have large
core radius ($r_c \stackrel{>}{\sim}0.5$ $kpc$), although there are
also
indications that $\sim 10\%$ of them have very small cores (FFKT and
references
therein). So, in view of the lack of more precise information we
assume that
$90\%$ of spiral galaxies have a constant core $r_c=0.5h^{-1}$$kpc$
and that
the remaining $10\%$ are well described by SIS. In fact these
assumptions are
enough to practically reduce the contribution of S galaxies to the
optical
depth  to only $10\%$ of its SIS value. Actually in case C
(cosmological
constant), remains another $\sim 1\%$ effect. We can understand this
small
difference by observing that the quantity
$D_{OL}D_{LS}/D_{OS}$ for fixed redshift is higher in case C (see
Figure $2$).
This means that the parameter $\beta$ tends to be smaller in case C
and
explains why the effect should be less important in this case. In any
case, we
confirm the conclusion obtained by Krauss and White, that spiral
galaxies have
a very small effect on lensing frequencies even if we do not include
the
$\sqrt{3/2}$ factor in the E/S0 dispersion velocities.

While spirals usually have large core radius, E and SO galaxies are
believed to
have smaller ones. Most of the analyses of elliptical  galaxies are
based on
slightly different relations between core radius and velocity
dispersion
(luminosity) that are usually derived by using  Lauer's (1985) study
of nearby
elliptical galaxies. Lauer found that 14 galaxies (from a total
number of 42)
have resolved cores
($80pc\stackrel{<}{\sim}r_c\stackrel{<}{\sim}400pc$), 23
were unresolved and 5 had marginally resolved cores.
By fitting Lauer's data for the resolved cores, Krauss and White
obtained a
relation between E core radius and luminosity. By assuming that
relation to be
valid for all E galaxies they obtained a suppression factor $\sim
0.4$ for the
Einstein-de Sitter model and $\sim 0.63$ for the cosmological
constant
dominated universe. Fukugita {\it et al.} (FFKT) assumed that $1/3$
of E
galaxies are well described by the relation they obtained from
Lauer's study.
They also assumed that another $1/3$ have core radius given by
multiplying that
relation by $1/3$ and that the remaining $1/3$ have $r_c \sim 10 pc$.
With this
model they obtained a suppression factor equal to $0.65$. They also
claim
having changed their assumptions in a reasonable way and always
getting numbers
between $0.5$ and $0.7$. We analysed this effect more quantitatively
and
reached similar results.

We also analysed the core radius effect in the case of empty beam and
ES-statistics. We found that the suppression is higher in this case.
The reason
is that $a_{cr}(0)$ for the empty beam case is smaller and this
implies that
$\beta$ is higher, thus increasing the suppression effect. Actually,
for spiral
galaxies the effect can be very high. For instance, from Figures $4a$
and $4b$
it is clear that $\tau_{SIS}$(z, filled beam)/$\tau_{SIS}$(z, empty
beam) is
less than one. This means that we should expect a higher frequency of
lensed
quasars in the open beam case.  However by looking at Figures $5a$
and $5b$ we
immediately realize that for the special choice of the parameters
$\tau_{NSIS}$(z, filled beam)/$\tau_{NSIS}$(z, empty beam) is higher
than one
and the opposite would be expected. In fact this occurred because we
considered
in our example a typical spiral galaxy with relatively high core
radius and
small velocity dispersion. For E/SO galaxies we should expect this
effect not
to be so conspicuous and, in fact, under reasonable assumptions we
obtained a
suppression factor that is only $\stackrel{<}{\sim} 10\%$ higher than
that in
the filled beam case. In the next section, when comparing model
predictions to
observations, we will take this small difference into consideration.
In order
to simplify calculations, we will follow FFKT and assume, for filled
beam
distance,  a constant core effect suppression factor equal to $0.65$
for E/SO
galaxies. In the case of the empty beam we will consider a
suppression factor
equal to $0.60$.

By using the filled beam distance, standard statistics and the SIS
profile it
can be shown (see FT and FFKT) that the normalized image angular
separation
distribution for a source at $z_S$  is given by
\begin{eqnarray}
\frac{dP}{d\theta}(z_S,\theta)  =  \frac1{\tau(z_S)}\int^{z_S}_0
\frac{d^2\tau}{dz_Ld\theta}dz_L \ \ \ \ \ \ \ \ \ \ \ \ \ \ \ \ \ \ \
\ \ \ \ \
\ \ \ \ \ \ \ \ \ \ \ \ \ \ \ \ \ \
\nonumber \\
=  \frac{f}{\tau(z_S)}\int^{z_S}_0
(1+z_L)^3\left(\frac{D_{OL}D_{ls}}{cH{_0}^{-1}D_{OS}}\right)^2
\left(-\frac 1{cH_{0}^{-1}}\frac{cdt}{dz_L}\right) \times
\nonumber \\
\frac{\gamma/2}{\Gamma(\alpha+1+\frac4\gamma)}
\left(\frac{D_{OS}}{D_{LS}}\frac{\theta }{ 8 \pi
\left(\frac{\sigma_{||}^{\star}}{c}\right)^2}
\right)^{\frac{\gamma}{2}(\alpha+2)}
\exp{\left[-\left(\frac{D_{OS}}{D_{LS}}\frac{\theta }{ 8 \pi
\left(\frac{\sigma_{||}^{\star}}{c}\right)^2}\right)^{\frac{\gamma}{2}
}
\right]}\frac{1}{\theta} dz_L.
\end{eqnarray}

In Figure $6a$ the predicted image separation distribution for a
source at
redshift $z_S=2.2$ (the quasar average redshift in the HST snapshot
survey) is
shown in two cases for E/SO galaxies. We took $\alpha=-1.1$, and to
simplify
the computation we considered the same velocity for E and SO
galaxies, that is
we chose an average velocity, $\sigma_{||}^{\star}=213 \times
\sqrt{\frac32}$
$km/s$ in case (i) and  $\sigma_{||}^{\star}=213$ $km/s$ in case
(ii). The same
quantity for both cases is also displayed in Figure $6b$ for the
empty beam
distance and Ehlers-Schneider statistics. In this case we took into
account
equation ($15$) in the definition of the differential optical depth.
Figure 6a
shows that for flat models and filled beam distance the image
separation
distribution is independent of $m$ and  $\Omega_{m0}$. In fact, as
remarked by
FFKT, it is also independent of $z_S$. For empty beam distance the
degeneracy
of flat models is broken and we can observe a shift of the
distribution to
smaller values of angular separation. It is also clear that in both
cases
increasing the velocity dispersion increases the probability of
larger image
separation.

\section{Comparison with observations}
\noindent

In this section we shall compare the predicted number of multiple
images (and
their angular image separation) for the models presented in section 2
with the
observational results of the Hubble Space Telescope Snapshot Survey
(Maoz {\it
et al.} 1993). In the last report of the survey, Maoz {\it et al.}
announced
the existence of six lens candidates from a sample of 502 quasars. In
fact, two
of the lensed candidates have unexpectedly large image separation.
There is
evidence that one of them was produced not by a single galaxy but by
a more
complex system, and it is not clear that the other one is really a
multiple
image of one single object (Maoz and Rix and references therein) .
Following
the current interpretation, we shall not  include these two cases in
our
analysis. The four remaining candidates have the following redshift
and image
separation ($z$,$\theta$): ($3.8$, $0.47"$), ($2.55$, $1.22"$),
($1.72$,
$2.0"$) and ($2.72$, $2.2"$).

The expected number of lensed quasars in the survey ($N_E$), for each
model, is
obtained by computing the quantity
\begin{equation}
N_E=\sum_{i=1}^{N_Q} \tau (z_i)
\end{equation}
where the sum is over the $N_Q$ quasars redshifts of the survey. In
calculating
the expected number of multiple images, besides the core radius
effect that we
discussed in the last section, we also took into account two other
corrections
to $\tau$, namely, the angular selection effect ( $\times 0.95$ for E
and SO)
and magnification bias ($\times 9.1$) (see Fukugita and Peebles).

In Table 1 we display, for each model, the predicted number of lensed
quasars
for the HST snapshot survey. The numbers were obtained for filled
beam and
standard statistics as well as for empty beam and Ehlers-Schneider
statistics.
We show results for the cases in which we have and have not included
the
$(3/2)^{1/2}$ factor in velocity dispersion. By assuming Poisson
statistics we
also  display, for all cases, the probability of detecting the
observed number
(four) of lensed quasars. Notice that in case C the final results
depend
considerably on the assumptions performed. For instance, if we take
into
account the $(3/2)^{1/2}$ factor, the predicted number of lensed
quasars is too
high in model C. However if we do not consider the correction the
model cannot
be ruled out, at least based only on the total number of images. From
Table 1
we also see that model D is the least sensitive to the assumptions
and seems to
be in good agreement with observations. It is also clear that the
lensing rate
is a factor $\sim 2$ smaller in model B than in model C and this
corroborates
the idea that usually decay $\Lambda$  models have a lower lensing
rate than
the constant $\Lambda$  ones. We should remark that this is not
always true and
some decaying $\Lambda$ models can, in fact, predict very large
lensing rates.

Recently Kochanek (1993) pointed out that analyses based only on
Poisson
statistics give no weight to the different ways that a fixed number
of multiple
images is produced. He then suggested a maximum likelihood method
that takes
this into account. His technique is based on the following likelihood
function,
\begin{equation}
 L=\prod _{i=1}^{N_U} (1-p_i) \prod _{j=1}^{N_L} p_j \prod
_{k=1}^{N_L} p_{ck}.
\end{equation}
Here $N_U$ is the number of unlensed quasars, $N_L$ is the number of
lensed
ones, $p_i\ll1$ is the probability that quasar $i$ is lensed and
$p_{ck}$ is
the configuration probability, that we shall consider as the
probability that
quasar $k$ is lensed with the observed image separation.

We applied Kochanek technique to the flat models. By expressing $L$
as a
function of the parameters $m$ and $\Omega_{m0}$ we obtained the
maximum of the
likelihood function ($L_{max}$) and formed the ratio $l=L/L_{max}$.
It can be
shown that with two parameters, the distribution of $-2\ln{l}$ tends
to a $\chi
^2$ distribution with two degrees of freedom (Kendall \& Stuart 1977,
Eadie
{\it et al.} 1971, Kochanek 1993).

Contours of constant likelihood are plotted in Figure $7$. Regions
with larger
likelihood are represented by lighter shades. The maximum of the
likelihood is
indicated by a cross ($+$). We should remark that the figure is very
broad and
is displayed to give a qualitative view of the likelihoods. For the
figure we
considered the value of the velocity dispersion equal to 213 km/s. We
observed
that the results did not change appreciably when we increased the
velocity
dispersion to $\sigma_{||}^{\star}=261$ km/s. Some qualitative
aspects can be
inferred from the figure. For instance, if $m$ is low
($m\stackrel{<}{\sim}
0.5$) regions with a lower value of $\Lambda$, let say, those with
$\Omega_{m0}\stackrel{>}{\sim}0.4$, have higher likelihood. However
if
$m\stackrel{>}{\sim} 2$ that constraint does not exist and models
with low
$\Omega_{m0}$ are in fact more likely. We observe that in the
two-dimensional
space of the parameters of Figure 7, the Einstein-de Sitter models is
represented by two points. The first one is $m=0$ and
$\Omega_{m0}=1$, while
the second one is obtained by taking the limit $m\rightarrow 3$ and
$\Omega_{m0}\rightarrow 0$.
Figure 8 gives more quantitative information. We plot in it the
$50\%$,
$68\%$($1\sigma$) and $95.4\%$($2\sigma$) confidence levels for the
likelihood
ratio for the two parameters.

It is important to emphasize that our main goal in this paper was not
to obtain
constraints on possible values of a constant $\Lambda$. Our target in
this work
was rather to show that if $\Lambda$ is not constant, the lensing
constraints
on $\Omega_{\Lambda}$ are much weaker. In fact we found that regions
with
higher values of $m$ and lower $\Omega_{m0}$ present larger
likelihood.
However, if we want to make some comparison with previous results, as
for
instance those of Kochanek (1993), we should fix the parameter $m$ to
the value
$m=0$ (cosmological constant). In this case the  likelihood peaks at
$\Omega_{m0}=0.62$  but now we have that $-2\ln{l}$ is distributed
like a $\chi
^2$ distribution with one degree of freedom. In this case the
constraints on
$\Lambda$ are stronger and, for instance, we get
$\Omega_{\Lambda}\stackrel{<}{\sim}0.8$ at $90\%$ confidence level.
We remark
that we have not obtained one of Kochanek's constraints,
$\Omega_{\Lambda}
\stackrel{<}{\sim} 0.45$, for the following reasons. First, as we
discussed, we
 were only considering lensing by galaxies, so we have not included
in our
analysis the lens 0957+561. Second, we have used the SIS profile but
we took
into account a core radius suppression effect. Third, in our
likelihood
analysis we neglected the contribution of spiral galaxies.

\section{Discussion and conclusion}
\noindent

As is well known, a sufficiently long period of inflation in the
early universe
gives a natural solution to the isotropy, flatness, and monopole
problems. A
nearly invariant primordial spectrum is also generated in most of the
inflationary models of the universe, a feature that seems to be in
good
agreement with observations. Inflationary models usually predict
$\Omega_{total}=1$. Nevertheless, observations on the scales $10-30$
$Mpc$,
based on dynamical methods, indicate $\Omega_{m0}=0.2\pm0.1$.
Further, in flat
models with $\Omega_{m0}=1$ only if $h\leq 0.59$ is it possible to
get
theoretical ages in agreement with the lowest age estimates ($t_0=11
Gyr$) for
globular clusters. However, most of the recent observations indicate
higher
values for the Hubble parameter. Besides, the standard CDM model
($\Omega_{m0}=1$ and $h=0.5$) when normalized to COBE  predicts more
power on
small scales than is observed, and some of its variants such as
$HCDM$ seem to
work well only if $h\sim0.5$ (Primack 1995). By assuming a non-zero
cosmological constant all these problems can be solved at once while
keeping
the attractiveness of inflation.

Cosmologists in general show an enormous resistance in accepting the
idea of a
non-zero cosmological constant for several reasons. First, because it
is
another parameter in the theory and from an aesthetic point of view
this makes
$\Lambda$-models less compelling. The second reason is that in order
to
dominate the dynamics of the universe only recently, this parameter
should have
a very small value ($\Lambda \stackrel{<}{\sim} 10^{-56} cm^{-2}$)
that is in
fact $50$ to $120$ orders of magnitude below the estimate given by
quantum
field theory. If we assume a decaying cosmological term, this second
problem is
alleviated but we have to pay the price of introducing another
parameter. In
this paper we have also discussed a weakness on the observational
side of
constant-$\Lambda$ models. It is the prediction of too high frequency
of lensed
quasars by models with a large cosmological constant. Ratra and
Quillen showed
that the predicted lensing rate is considerably reduced in certain
models in
which a scalar field plays the role of an effective decaying
$\Lambda$-term. In
this paper we reached the same conclusion for models in which the
cosmological
term decreases with time as $\Lambda \propto a^{-m}$. We went one
step further
and also showed that for these models, lower values of $\Omega_{m0}$
and larger
 $m$ have higher likelihood. Finally, we should  mention that by
increasing the
value of the parameter $m$, and maintaining constant the value of
$\Omega_{m0}$, the theoretical age predicted by the models we have
considered
decreases. So, it would be interesting to extend the likelihood
analysis by
also taking into account the age constraints. Further investigation
in this
direction is being carried out.
\begin{center}
{\bf Acknowledgements}
\end{center}
We would like to thank Josh Frieman for critically reading the
manuscript and
for several useful suggestions.
We also would like to thank Luca Amendola, and Scott Dodelson for
helpful
discussions and Edgar Oliveira for calling our attention to a useful
reference
on statistics. This work was supported in part by the Brazilian
agency CNPq and
by the DOE and NASA at Fermilab through grant NAG5-2788.

\begin{center}
{\bf References}
\end{center}
\noindent
Abdel-Rahman, A-M. M.: 1992, {\sl Phys.Rev. \/} {\bf D45}, 3497. \\
Blandford, R.D., Kochanek, C.S.: 1987, {\sl `Dark matter in the
Universe',
World Scientific, Singapore,\/} p. 133, ``Gravitational lenses'',
edited by
J. Bahcall, T. Piran and S. Weinberg \\
Bourassa, R. R., Kantowski, R.: 1975, {\sl Astrophys. J.\/} {\bf
195}, 13.\\
Carrol, S. M., Press, W. H., Turner,
E. L.: 1992, {\sl  Annual Review of Astron. Astrophys.\/} {\bf  30},
499. \\
Carvalho, J. C., Lima, J. A. S., Waga, I.: 1992, {\sl Phys. Rev.
\/}{\bf D46},
2404. \\
Chaboyer, B.: 1995, ``Absolute ages of globular clusters and the age
of the
Universe'', submitted to ApJ {\em Letters\/}, astro-ph 9412015. \\
Chen. W., Wu, Y. S.: 1990, {\sl Phys. Rev \/} {\bf D41}, 695. \\
Dyer, C.C., Roeder, R.C.: 1972, {\sl Astrophys. J.\/}
{\bf 174}, L115. \\
Dyer, C. C., Roeder, R. C.: 1973, {\sl Astrophys. J.\/} {\bf 180},
L31.\\
Eadie, W. T., {\it et al.}: 1971, ``{\it Statistical methods in
experimental
physics}'', North-Holland Publishing Company. \\
Efstathiou, G., Sutherland, W. J. \& Maddox, S. J.: 1990, {\sl
Nature\/}
{\bf 348}, 705. \\
Ehlers J. and Schneider P.:1986, {\sl Astron. Astrophys.\/} {\bf
168}, 57.
\\
Faber, S., Jackson, R.: 1976, {\sl Astrophys. J.\/} {\bf 204}, 668.
\\
Feldman, H. A., Evrard, A. E.: 1993, {\sl Int. J. Mod. Phys.\/} {\bf
D2}, 113.
\\
Freedman, W. L., {\it et al.}: 1994, {\sl Nature\/} {\bf 371}, 757.
\\
Freese, K., Adams, F. C., Frieman, J. A., Mottola, E.: 1987, {\sl
Nucl.
Phys.\/} {\bf B287}, 797. \\
Fry, J. N.: 1985, {\sl Phys. Lett.\/} {\bf B158}, 211. \\
Fukugita M., Futamase T., Kasai M.: 1990, {\sl Monthly Notices Roy.
Astron.
Soc.\/} {\bf 246}, 24p. \\
Fukugita M., Futamase T., Kasai M., E. L. Turner: 1992, {\sl
Astrophys. J.\/}
{\bf 393}, 3. \\
Fukugita, M., Peebles, P.J.E.: 1993, {\sl Astrophys. J.,
submitted\/}, ``Visibility of gravitational lenses and the
cosmological
constant problem'' \\
Hinshaw, G.: 1988, ``{\it Topics in gravitational theory}'', PhD
thesis,
Harvard University. \\
Hinshaw, G., Krauss, L.M.: 1987, {\sl Astrophys. J.\/}
{\bf 320}, 468.\\
Kendall, M., Stuart, A.: 1977, ``{\it The advanced theory of
statistics}'',
Volume 2, $4^{th}$ edition, Charles Griffen, London. \\
Kochanek, C.S.:  1993, {\sl  Astrophys. J.\/} {\bf 419},  12. \\
Kofman, L. A., Gnedin, N. Y., Bahcall, N. A.: 1993, {\sl Astrophys.
J.\/}
{\bf 413}, 1. \\
Krauss, L., White, M.: 1992, {\sl Astrophys. J.\/} {\bf 394
}, 385.\\
Lahav, O., Lilje, P. B., Primack, J. R.\& Rees, M. J.: 1991, {\sl
Monthly
Notices Roy. Astron. Soc.\/} {\bf 251}, 128.\\
Lodge, O. J.: 1919, {\sl Nature\/} {\bf 104}, 354. \\
Mao, S.: 1991, {\sl Astrophys. J.\/} {\bf 380}, 9. \\
Mao, S., Kochanek, C.S.: 1994, {\sl Monthly Notices Roy. Astron.
Soc.,
submitted\/}, ``Limits on galaxy evolution''\\
Maoz, D., {\it et al. }: 1993, {\sl  Astrophys. J.\/} {\bf  409}, 28.
\\
Maoz, D., Rix, H. W.:  1993, {\sl  Astrophys. J.\/} {\bf  416},  425.
\\
Martel, H.: 1995, ``Nonlinear structure formation in flat
cosmological
models'', to appear in {\sl Astrophys. J.\/}, University of Texas
McDonald
Observatory preprint No. 305. \\
Ozer, M., Taha, M. O.: 1987a, {\sl Phys. Lett.\/} {\bf B171}, 363.
\\
Ozer, M., Taha, M. O.: 1987b, {\sl Nucl. Phys.\/} {\bf B287}, 776.
\\
Peebles P. J. E.: 1984, {\sl Astrophys. J.\/} {\bf 284}, 439.\\
Peebles P. J. E.: 1993, ``{\it Principles of physical cosmology}'',
Princeton
University Press.\\
Peebles,P. J. E., Ratra, B.: 1988, {\sl Astrophys. J.\/} {\bf 325
}, L17.\\
Pierce, M. J., {\it et al.}: 1994, {\sl Nature\/} {\bf 371}, 385. \\
Press, W.H., Gunn, J.E.: 1973, {\sl Astrophys. J.\/} {\bf 185},
397.\\
Primack, J. R.: 1995, {\it Status of cosmological parameters: Can
$\Omega=1$?},
to appear in the ``Proceedings of Particle and Astrophysics in the
Next
Millenium'', eds. E. W. Kolb and R. Peccei, World Scientific. \\
Ratra, B., Peebles, P. J.E.: 1988. {\sl Phys. Rev.\/} {\bf D37},
3407. \\
Ratra B., Quillen A.: 1992, {\sl Monthly Notices Roy. Astron. Soc.\/}
{\bf
259}, 738.\\
Refsdal S.: 1964, {\sl Monthly Notices Roy. Astron. Soc.\/} {\bf
128}, 307. \\
Rix, H.-W., Maoz, D., Turner, E.L., Fukugita,M.: 1994,
{\sl Astrophys. J., submitted\/}, ``Galaxy mergers and gravitational
lens
statistics''\\
Sahni, V., Feldman, H. A., Stebbins, A.: 1992, {\sl  Astrophys.
J.,\/} {\bf
385}, 1. \\
Schneider P., Ehlers J., Falco E. E.: 1992, ``{\it Gravitational
Lenses}'', Springer-Verlag. \\
Silveira V., Waga I.: 1994, {\sl Phys. Rev \/} {\bf D50}, 4890.\\
Subramanian, K., Cowling, S.A.:  1986, {\sl  Monthly Notices Roy.
Astron.
Soc.\/} {\bf  219},  333. \\
Tully, R. B., Fisher, J.: 1977, {\sl  Astron. Astrophys.\/}, {\bf
54}, 661. \\
Turner, E.L.: 1990, {\sl Astrophys. J.\/} {\bf 365}, L43. \\
Turner E. L., Ostriker J. P. , Gott III J. R.: 1984, {\sl Astrophys.
J.\/} {\bf
284}, 1, (1984).\\
Turner M. S., Steigman G. \& Krauss L. M.: 1984, {\sl Phys.
Rev. Lett. \/} {\bf 52}, 2090. \\
Vilenkin A.: 1984, {\sl Phys. Rev. Lett. \/} {\bf 53}, 1016,
(1984).\\
Young, P., Gunn, J.E., Kristian, J., Oke, J.B.,
Westphal, J.A.: 1980, {\sl Astrophys. J.\/} {\bf 241}, 507 \\
\newpage
\begin{table}
\caption{\bf Predicted number of lensed quasars for the HSTSS and
model
probabilities}
\begin{tabular}{ccccc}
 &model A&model B&model C&model D \\
 &$k=0$&$k=0, m=2$&$k=0, m=0$&$k=-1$ \\
&$\Omega_{m0}=1$&$ \Omega_{m0}=0.2$&$\Omega_{m0}=0.2$&$
\Omega_{m0}=0.2$ \\
\hline \hline \\
filled beam (with $\sqrt{3/2}$)&$2.6$&$4.7$&$10.7$&$4.1$ \\
probability&$18\%$&$17\%$&$0.3\%$&$19\%$ \\ \\
empty beam (with $\sqrt{3/2}$)&$4.2$&$7.1$&$14$&$4.6$ \\
probability&$19\%$&$6\%$&$ 10^{-2}\%$&$18\%$ \\ \\
filled beam (without $\sqrt{3/2}$)&$1.2$&$2.2$&$5.0$&$1.9$ \\
probability&$9\%$&$16\%$&$16\%$&$14\%$ \\ \\
empty beam (without $\sqrt{3/2})$&$1.9$&$3.3$&$6.8$&$2.1 $ \\
probability&$14\%$&$20\%$&$7\%$&$15\%$ \\ \\
\end{tabular}
\label{Table 1}
\end{table}
\clearpage
\begin{center}
{\Large {\bf Figure Captions} }
\end{center}

\bigskip
\noindent
{\bf Figure 1:} Dyer-Roeder distance as a function of the redshift;
(a) filled
beam and (b) empty beam.

\bigskip

\noindent
{\bf Figure 2:} The combination $D_{OL}D_{LS}/(D_{OS}cH_0^{-1})$ as a
function
of $z_L$ ; (a) filled beam distance and (b) empty beam distance.

\bigskip

\noindent
{\bf Figure 3:} Lensing geometry.

\bigskip

\noindent
{\bf Figure 4:} The normalized optical depth ($\tau /f$) as a
function of the
redshift for the SIS case; (a) is for filled beam distance and
standard
statistics and (b) for empty beam distance and Ehlers-Schneider
statistics.

\bigskip

\noindent
{\bf Figure 5:} The normalized optical depth ($\tau /f$) as a
function of the
redshift for the NSIS case; (a) is for filled beam distance and
standard
statistics and (b) for empty beam distance and Ehlers-Schneider
statistics. For
all the plots we assumed $r_c=0.5$$h^{-1} kpc$ and
$\sigma_{||}^\star=144$
$km/s$.

\bigskip

\noindent
{\bf Figure 6:} Distribution of the angular image separation for a
source at
$z_S=2.2$ in SIS case. We took $\alpha=-1.1$, $\gamma=4$ and
considered in case
(i) $\sigma_{||}^{\star}=261$ $km/s$ and in (ii)
$\sigma_{||}^{\star}=213$
$km/s$ ; (a) filled beam and standard statistics and (b) empty beam
and
Ehlers-Schneider statistics.

\bigskip

\noindent
{\bf Figure 7:} Contours of constant likelihood for flat models are
plotted in
the $\Omega_{m0}$ and $m$ parameter space. Regions with larger
likelihood are
represented by lighter shades. The peak of the likelihood is
indicated by the
cross.

\bigskip

\noindent
{\bf Figure 8:} Likelihood contours at $50\%$, $68\%$($1\sigma$) and
$95.4\%$($2\sigma$) confidence levels for the two dimensional
likelihood
distribution $l$.

\end{document}